\documentclass[useAMs,usenatbib,a4paper]{mn2e}
\usepackage{savesym}
\usepackage{graphicx}
\expandafter\let\csname equation*\endcsname\relax
  \expandafter\let\csname endequation*\endcsname\relax 
\usepackage{subfig}
\usepackage{amsmath}
\usepackage{amssymb}
\usepackage{verbatim}
\usepackage{array}
\usepackage{times}
\usepackage[total={17.8cm,24.0cm},centering]{geometry} 

\newcommand{\be}{\begin{equation}}
\newcommand{\beq}{\begin{equation}}
\newcommand{\ee}{\end{equation}}
\newcommand{\eeq}{\end{equation}}
\newcommand{\eea}{\end{eqnarray}}
\newcommand{\bea}{\begin{eqnarray}}

\newcommand\bb[1] { \mbox{\boldmath{$#1$}} }
\newcommand\bcdot{\bb{\cdot}}
\newcommand\del{\bb{\nabla}}

\def\gtsima{$\; \buildrel > \over \sim \;$}
\def\gtsim{\lower.5ex\hbox{\gtsima}}

\def\ltsima{$\; \buildrel < \over \sim \;$}
\def\ltsim{\lower.5ex\hbox{\ltsima}}

\title[Solar rotation and radiative equilibrium]{Differential rotation and radiative equilibrium in the Sun: is the tachocline spreading?}
\author[Andrea Caleo, Steven A. Balbus and William J. Potter]{Andrea Caleo \thanks{E-mail:
andrea.caleo@astro.ox.ac.uk}, Steven A. Balbus and William J. Potter
\\
Oxford Astrophysics. Denys Wilkinson Building, Keble Road, Oxford, OX1 3RH, United Kingdom}
\begin{document}
\date{}

\pagerange{\pageref{firstpage}--\pageref{lastpage}} \pubyear{2014}

\maketitle

\label{firstpage}

\begin{abstract}
It is well known that the combination of barotropic rotation and radiative equilibrium are mutually incompatible in stars.  The Sun's internal rotation is far from barotropic, however, which allows at least the theoretical possibility that the Sun's thermal balance is one of radiative equilibrium in the region of the tachocline near the outer boundary of the radiative zone.  We show here that (i) the constraint of radiative equilibrium leads to a
straightforward ordinary differential equation for the Sun's rotation profile, and (ii) solutions of this equation can be found that, to within current levels of accuracy, closely resemble the rotation profile deduced from helioseismology.  More generally, we calculate how large a baroclinic deviation from uniform rotation is required to maintain radiative equilibrium without meridional circulation throughout the bulk of the radiative zone.  Very little deviation is required, well below detectability. The feasibility of radiative equilibrium for the tachocline suggests that the issue of a spreading boundary layer may be less severe than previously thought.

\end{abstract}


\section{Introduction} \label{sec:data}

Since the availability of helioseismological data has allowed the reconstruction of the details of the internal rotation of the Sun, significant effort has been directed towards understanding what sustains the observed rotation rate $\Omega(r, \theta)$. The shape of the iso-rotation curves in the bulk of the convective zone was found to be well described by the characteristic solutions of the thermal wind equation \citep{BalbusBonartLatterWeiss2009}. In contrast, the physics of the relatively thin transition between the radiative and convective zone, known as the solar \emph{tachocline}, is still largely uncertain.

The well known theorem of Von Zeipel states that a star in uniform rotation cannot be in radiative equilibrium (see e.g. \citet{Schwarzschild1956}).  This is generally applied well within radiative cores, where any resdidual thermal energy imbalance is thought to be compensated by means of a mean velocity flow known as Eddington-Sweet circulation. However, \citet{BalbusSchaan2012} have noted that matters are not so simple near the outer boundary of the radiative zone in a Sun-like star.  The steady entropy equation, allowing for advection and radiative diffusion, is:
\begin{equation} \label{eqEntro}
P (\bb{u} \cdot \del) \sigma = - (\gamma - 1) \del \bcdot \bb{F},
\end{equation}
where $P$ is the pressure, $\bb u$ is the circulation velocity, $\sigma = \log{(P \rho^{- \gamma})}$ is the entropy variable, $\gamma$ is the adiabatic index and $\bb F$ is the radiative flux. The radial, dominating component of $\del \sigma$ is positive in the radiative zone and negative in the convective zone as for the Schwarzschild criterion, so that it has to go through zero at the radiative-convective boundary. Therefore, there are two possibilities at the boundary surface: either the circulation velocity is high enough for the $(\bb u \cdot \del) \sigma$ term to be relevant and the angular component of $\del \sigma$ is finite, or the $(\bb u \cdot \del) \sigma$ term is negliglible and equation \eqref{eqEntro} reduces to $\del \cdot \bb F = 0$, the condition of radiative equilibrium.  Neither of these possibilities is consistent with uniform rotation (or barotropic rotation more generally), and indeed the Sun shows a strong departure from rotation on cylinders in this region (and elsewhere).   

The Eddington-Sweet circulation velocity in the radiative zone of stars has a prominent role in the literature for solving the problem of energy transfer under conditions of uniform rotation. However, the inclusion of meridional circulation in the equations of stellar structure generates further complications (see \citet{Tassoul2000} for a discussion). The problem originates from the time-steady azimuthal component of the Euler equation of motion:
\begin{equation} \label{Euler0}
(\bb v \cdot \del) \bb v = -\frac{1}{\rho} \del P + \bb g,
\end{equation}
that, for $\bb v = \Omega R \hat \phi + \bb u$, where $\Omega$ is the angular velocity, $R$ the cylindrical radius and $\hat \phi$ the azimuthal direction, gives
\begin{equation} \label{AngMom0}
\bb u \cdot \del (\Omega R^2) = 0,
\end{equation}
i.e. for a star in uniform rotation, the circulation is along cylindrical surfaces at fixed distance from the rotation axis. This is incompatible with a model that is steady and symmetric with respect to the equatorial plane, as material would accumulate onto such a plane due to the circulation velocity.   One is then compelled to consider non-steady models or to introduce an accommodating magnetic field structure that would change equation \eqref{AngMom0}.   Building a simple, comprehensive model for rotating stars when meridional circulation is included is not straightforward.

For these reasons, it is interesting to ask whether the current data on the rotation of the outer part of the radiative zone of the Sun are compatible with a model in \emph{strict radiative equilibrium.}   Even in the bulk of the radiative zone, which is in near uniform rotation, the current accuracy of the data certainly does not allow one to say that it is rotating \emph{exactly} as a rigid body.  A small amount of differential rotation could be present.  Would this undetected deviation from uniformity be enough for the Sun to be in radiative equilibrium?

The accuracy of the data of helioseismology decreases with depth in the Sun, and it is particularly difficult to study the properties of a narrow transition layer like the tachocline, whose thickness may appear to be broader than it actually is because of the resolution of the inversion techniques involved in the data analysis.  There are several different estimates of the width of the tachocline (see \citet{Christensen2007}), all in the range $1\% - 5 \% $ R$_\odot$; the most recent analysis by \citet{Antia2011} suggests the possibility of a significant dependence of the tachocline width on solar latitude, with an average value not larger than $3 \%$ R$_\odot$ but a strong uncertainty for the value at mid-latitudes (see their figure 2). This is thinner than the transition apparent from the data used in this paper. It is likely that the structure of the tachocline is still substantially unresolved.   Any direct comparison of theory with the data must therefore be considered as a test of the plausibility of the model, not as a precision attempt to reproduce the actual $\Omega(R, z)$ in the interior of the Sun.  The uncertainties are too significant.

A discussion of the current knowledge of the internal rotation of the Sun through the radiative zone to the core is given in \citet{Howe2009}. Estimating the systematic error in the rotation data for the interior of the Sun is not easy because of the complex inversion procedure involved \citep{Chaplin1999}. It is currently accepted that the data for the bulk of the convective zone of the Sun are very accurate, while those for the interior core are very uncertain.   Although a thorough estimate of the errors in the outer radiative zone is not available, it is likely to be intermediate between the convective zone and the core, with a relative uncertainty that is perhaps of order $10\%$; if one is conservative, possibly more (R. Howe, private communication).

With these caveats in mind, in this paper, we construct a model of the region of the Sun interior to the convective zone under the assumption of radiative equilibrium, use this constraint to compute the rotation profile itself, and compare the resulting angular velocity curves with the observed differential rotation of the Sun.  Significantly, we find that this model is compatible with the current data.

\subsection{The spreading of the tachocline}      \label{secSpreading}
An influential analysis of the tachocline was conducted by \citet{SpiegelZahn1992} (see also \citealt{Zahn2007}) at a time when the early results of helioseismology showed that the rotation of the Sun turned from strongly differential to uniform in a thin, unresolved transition at the radiative - convective boundary. These authors considered the standard equations of stellar structure, including the effect of circulation velocity.   A time-dependent, initial-value problem with boundary conditions at the tachocline emerged. In the Spiegel \& Zahn calculation, the term that offset the advection of angular momentum in equation \eqref{AngMom0} above was $\partial \Omega / \partial t$, the time-explicit inertial derivative in the azimuthal equation of motion.    Time-dependence thus played an essential role in the analysis from the start.

The inclusion of this term led to the conclusion that the tachocline transition must spread with time.   Specifically, \citet{SpiegelZahn1992} showed that under these assumptions, a time-dependent diffusion-like equation for the angular velocity followed, implying ongoing penetration of the differential rotation into the radiative core of the Sun on a time-scale of $\tau \sim 10^9$ yr, significantly shorter than the age of the Sun.  This posed a fundamental problem: why is the tachocline so thin? In the last few years, a significant amount of theoretical work has been devoted to studying candidate mechanisms that would confine the spreading of an infinitesimally thin tachocline, both purely hydrodynamical (reviewed in \citealt{SolarTachocline2007}) and magnetohydrodynamical (\citealt{GoughMcIntyre1998}, \citealt{Garaud2007}).    However, the helioseismology data have not improved enough either to resolve the tachocline or determine a lower boundary for its thickness, or to significantly constrain its properties.

The differential rotation of the Sun has been examined by numerical hydrodynamic and magnetohydrodynamic solar models, and important results have been obtained in reproducing rotation profiles for the convective zone that qualitatively resemble the results of helioseismology.   \cite{Miesch2006} imposed ad-hoc entropy boundary conditions at the base of the convective zone, and more recently there have been 2D \citep{Rogers2011} and 3D \citep{Brun2011} simulations of the coupled convective and radiative regions. The simulations are limited by the necessity of including artificially large diffusivities, which prevent an accurate force balance in the interior and in the tachocline, and coping with a wide range of time-scales, from those of convection and internal gravity waves to the predicted long time-scale of tachocline spreading.  The results of \citet{Rogers2011} (who finds that hydrodynamic processes can significantly slow down the spreading of the tachocline) and \citet{Brun2011} (who find a very fast spreading of the tachocline but relate it to a large viscous diffusivity), while significant on their own terms, may not fully capture the true dynamics of the radiative-convective boundary.

The result by \citet{SpiegelZahn1992} implies that a thin (nearly discontinuous) tachocline, with initial and boundary conditions for the base of the convective zone drawn from helioseismology, is not a steady solution for the Sun, but instead gives rise to ``burrowing'' by differential rotation. This does not rule out the existence of a model of a steady, thin tachocline with a self-consistent angular velocity differing from the \citet{SpiegelZahn1992} profile,  but nevertheless compatible with the present helioseismology data. In this paper, we present such a model.

An outline of the paper is as follows.   In \S 2 we present a detailed analysis of the governing equations, arriving at a set of two coupled ordinary differential equations that, with suitable boundary conditions, allow the angular velocity to be reconstructed.   In \S 3, we apply our results to
the solar tachocline and radiative interior.    Finally, in \S 4 we summarize our conclusions.

\section {Description of the model}
\subsection{Equations of stellar equilibrium} \label{secPertur}
The steady equations of stellar equilibrium in the outer radiative zone for an inviscid flow are given by:
\begin{equation} \label{Euler}
(\bb v \cdot \del) \bb v = -\frac{1}{\rho} \del P + \bb g,
\end{equation}
\begin{equation} \label{EOS}
P = \frac{\rho}{\mu m_P} k_B T,
\end{equation}
\begin{equation} \label{DivF}
\del \cdot \bb F_{rad} = 0,
\end{equation}
\begin{equation} \label{FRad}
\bb F_{rad} = - \frac{4 a c T^3}{3 \rho \kappa} \del T,
\end{equation}
where $\bb{v}$ is the rotation velocity, $\rho$ the density, $P$ the gas pressure, $\bb{g}$ the (self) gravitational field, $m_p$ the atomic mass unit, $\mu$ the mean molecular mass, $T$ the temperature, $\bb{F}_{rad}$ the radiative flux, $a$ the radiation constant, $c$ the speed of light, and $\kappa$ the Rosseland mean opacity. 
For a given chemical composition $\kappa$ is a function of the local density and temperature: $\kappa = \kappa(\rho, T)$.

We denote the rotational velocity in spherical coordinates  $(r, \theta, \phi)$ by $\bb v = \Omega r \sin{\theta} \hat \phi$, introducing the angular velocity $\Omega$.   The rotation is considered to be a small perturbation to a static spherical structure, so that the equations can be linearized in the Eulerian perturbation variables $\delta P$, $\delta \rho$, $\delta \bb g$.   
\begin{equation} \label{Euler1}
\Omega^2 r \sin^2{\theta} - \frac{1}{\rho} \frac{\partial \delta P}{\partial r} - \frac{g}{\rho} \delta \rho + \delta g_r = 0,
\end{equation}
\begin{equation} \label{Euler2}
\Omega^2 r \sin{\theta} \cos{\theta} - \frac{1}{r \rho} \frac{\partial \delta P}{\partial \theta} + \delta g_\theta = 0,
\end{equation}
\begin{equation} \label{Radiative1}
\del \cdot \Big( \frac{ac}{3 \rho \kappa} \Big( \del (4 T^3 \delta T) - \Big( \frac{\delta \kappa}{\kappa} + \frac{\delta \rho}{\rho} \Big) \del T^4 \Big) \Big)  = 0,
\end{equation}
\begin{equation} \label{EOS1}
\frac{\delta T}{T} = \frac{\delta P}{P} - \frac{\delta \rho}{\rho},
\end{equation}
where the first two equations are the $\bb{\hat r}$ and $\bb{\hat \theta}$ components of the Euler force balance equation.  The unsubscripted $\rho$, $P$ and
$T$ quantities refer now to the unperturbed, nonrotating spherical equilibrium solution, and the perturbed Rosseland opacity $\delta \kappa$ is given by:
\begin{equation} \label{}
\delta \kappa = \Big( \frac{\partial \kappa}{\partial \rho}\Big)_T \delta \rho + \Big( \frac{\partial \kappa}{\partial T}\Big)_\rho \delta T.
\end{equation}
In the following analysis, we will neglect the self-gravity perturbations $\delta g_r$, $\delta g_\theta$ in equations \eqref{Euler1} and \eqref{Euler2}, following the Cowling approximation.   The validity of thie approximation is discussed in in Appendix A.

\subsection{Expansion in $\bb{\cos\theta}$}
\begin{figure*}
	\centering
		 \includegraphics[width=\textwidth, clip=true, trim=0cm 0cm 0cm 0cm]{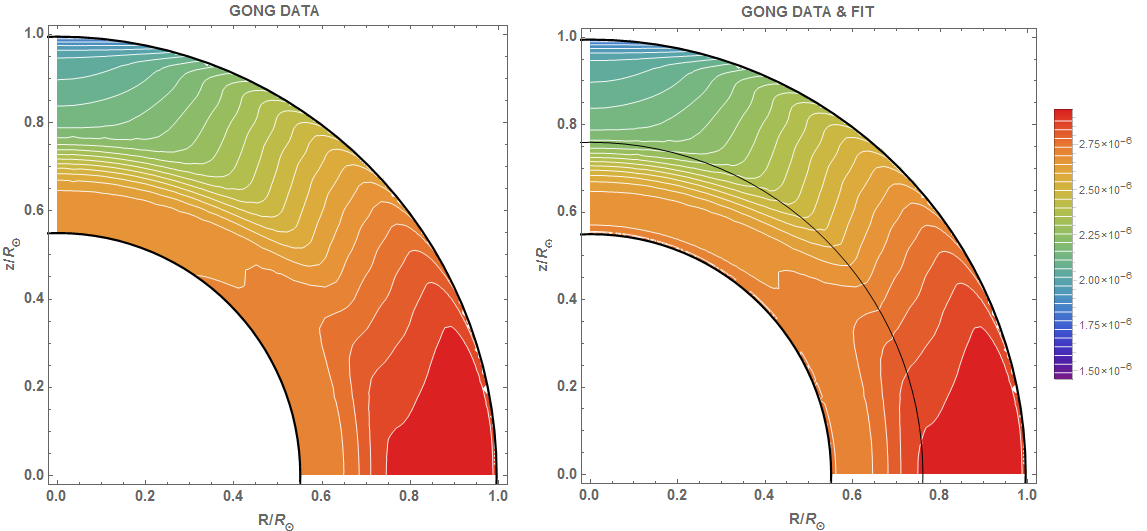} 
	\caption{\label{figOmegaDataFit}Isocontours of the angular velocity $\Omega(r, \theta)$ for $0.55 R_\odot < r < R_\odot$. The left panel shows the GONG data. The right panel shows our best fit of the form \eqref{OmegaFit} for the data interior to $r < 0.76 R_\odot$, joined with the actual GONG data for $r > 0.76 R_\odot$ (the surface $r = 0.76 R_\odot$ is shown as a black line).  The scale is in rad s$^{-1}$.  Figure \ref{figOmegaDataFitbis} shows a more detailed comparison of the values in the interval $0.55 R_\odot < r < 0.76 R_\odot$.}
\end{figure*}

\begin{figure*}
	\centering
		 \includegraphics[width=\textwidth, clip=true, trim=0cm 0cm 0cm 0cm]{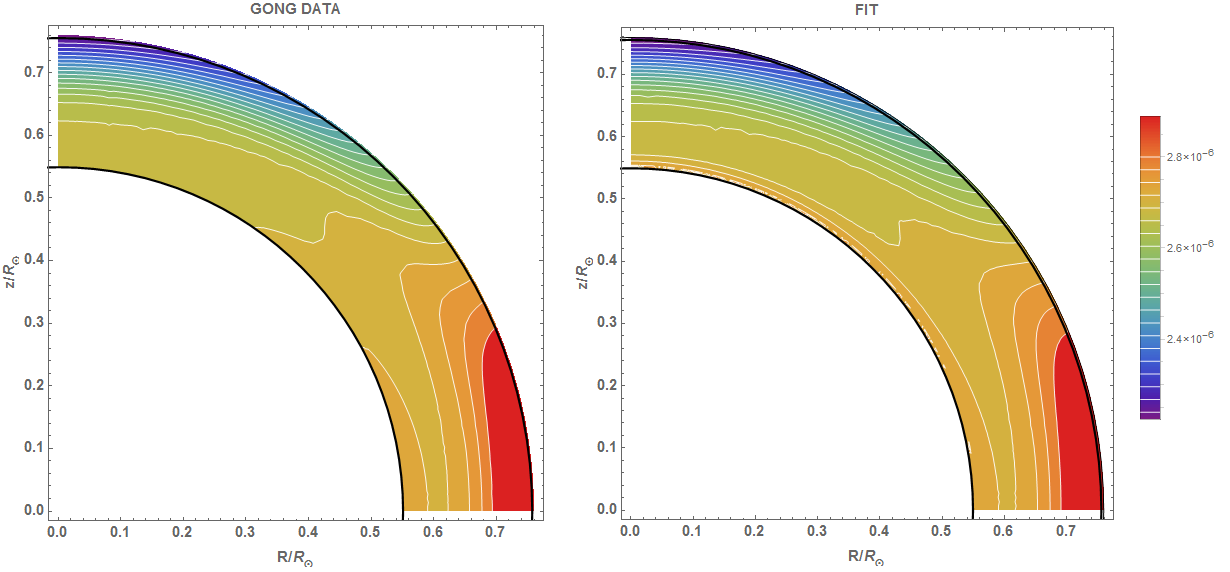} 
	\caption{\label{figOmegaDataFitbis}Isocontours of the angular velocity $\Omega(r, \theta)$ for $0.55 R_\odot < r < 0.76 R_\odot$. As in figure \ref{figOmegaDataFit}, the left panel shows the GONG data, while the right panel shows our best fit of the form \eqref{OmegaFit} for the data. The scale is in rad s$^{-1}$.}
\end{figure*}

To solve equations \eqref{Euler1}-\eqref{EOS1} in the variables $\Omega^2$, $\delta \rho$, $\delta P$ and $\delta T$, it is convenient to expand these quantities in even powers of $\cos{\theta}$.  The solar angular velocity given by the GONG data \citep{GONGData} in the outer part of the radiative zone is very well approximated by a function of the form:
\begin{equation} \label{OmegaFit}
\Omega^2(r, \theta) \cong \Omega_0^2(r) + \Omega_2^2(r) \cos^2{\theta}.
\end{equation}
The accuracy of this fit may be seen in figures \ref{figOmegaDataFit} and \ref{figOmegaDataFitbis}, which show the angular velocity isocontours of the Sun according to the GONG data (left) and an expansion of the form \eqref{OmegaFit} for $r < 0.76 R_\odot$ in which $\Omega_0^2(r)$, $\Omega_2^2(r)$ are 5-th order polynomials in $r$ optimised to fit the data (right). The functions $\Omega_0^2(r)$ and $\Omega_2^2(r)$ are shown as the solid lines in figures \ref{figOmega0}, \ref{figOmega2}. We have used an expansion of the type \eqref{OmegaFit} for the model of the radiatively sustained angular velocity curves.

It has been noted by \citet{BalbusSchaan2012} (see their equation (85)) that if $\Omega^2(r, \theta)$ can be expressed in terms of the first $n$ powers of $\cos^2{\theta}$, the equations of equilibrium can only be satisfied if the perturbations of the structural variables $\delta P$ and $\delta \rho$ have terms up to $n+2$ in their expansion:
\begin{equation} \label{PExp}
\delta P(r, \theta) = \sum_{n=0,2,4} P_n(r)\cos^n\theta,
\end{equation}
\begin{equation} \label{rhoExp}
\delta \rho(r, \theta) =\sum_{n=0,2,4} \rho_n(r)\cos^n\theta.
\end{equation}

\subsection{Solutions of the model in radiative equilibrium}
Expressions \eqref{OmegaFit}, \eqref{PExp} and \eqref{rhoExp} can be substituted into equations \eqref{Euler1} to \eqref{Radiative1} (in which $\delta T$ is expressed in terms of $\delta P$, $\delta \rho$ by means of \eqref{EOS1}) and the sum of all the terms proportional to the same power of $\cos{\theta}$ must be zero. Equation \eqref{Euler2} then yields for $n \geq 2$:
\begin{equation} \label{Pn}
P_n(r) =- \frac{1}{n} r^2 \rho \Omega_{n-2}^2(r),
\end{equation}
and equation \eqref{Euler1} gives, again for $n \geq 2$:
\begin{equation} \label{rhon}
\rho_n(r) = \frac{1}{g} \left(\rho r (\Omega_n^2(r) - \Omega_{n-2}^2(r)) - \frac{d P_n(r)}{dr}\right).
\end{equation}
The perturbations in the pressure and density can therefore be expressed in terms of the functions $\Omega_n^2(r)$ from the expansion of $\Omega^2(r, \theta)$, except for the spherically symmetric terms $P_0(r)$, $\rho_0(r)$.  Knowledge of those terms is not required to impose radiative equilibrium, and they can ultimately be considered as absorbed into the non rotating, spherically symmetric solution.

We may eliminate $\Omega^2_n$ from the above, obtaining a relation between $\rho_n$ and $P_n$:
\beq\label{rhobis}
r\rho_n = \frac{1}{r}\, [ nP_n - (n+2) P_{n+2}] - \frac{dP_n}{dr}.
\eeq
The same technique applied to the radiative equilibrium equation \eqref{Radiative1} leads to the rather cumbersome form
\begin{equation} \label{Radiative2}
	\begin{aligned}
	&\frac{5 C}{\rho r^2} \frac{d}{dr} \Big( \frac{\rho_n}{\rho} \Big)  - \frac{4 C}{P r^2} \frac{d}{dr} \Big( \frac{P_n}{P} \Big) + \frac{1}{r^2} \frac{d}{dr} \Big\{ \frac{r^2 T^4}{\rho \kappa}  \Big[ \Big( \frac{P_n}{P} \Big)' - \Big( \frac{\rho_n}{\rho} \Big)' \Big] \Big\}+ & \\
	&+   \frac{C}{r^2} \frac{d}{dr} \Big\{ \frac{1}{\kappa} \Big[ \rho_n \frac{\partial \kappa}{\partial \rho} + T \frac{\partial \kappa}{\partial T} \Big( \frac{P_n}{P} - \frac{\rho_n}{\rho} \Big) \Big] \Big\} + &\\ 
	&+ \frac{T^4}{r^2 \rho \kappa} (n+1) \Big[ (n+2) \Big( \frac{P_{n+2}}{P} - \frac{\rho_{n+2}}{\rho} \Big) - n \Big( \frac{P_n}{P} - \frac{\rho_n}{\rho} \Big)  \Big] = & \\
	& = 0,
	\end{aligned}
\end{equation}
where the constant
\begin{equation}
C = \frac{3 L_\odot}{16 \pi a c},
\end{equation}
has been introduced, where $L_\odot$ is the Solar luminosity, and we have used prime notation $(X)'$ for $dX/dr$ to aid readability.  This equation holds for $n=2, 4$, with the understanding that $P_6 = 0$, $\rho_6 = 0$.   Equations \eqref{rhobis} and \eqref{Radiative2}, with proper boundary conditions, are coupled ordinary differential equations that uniquely determine the solution of our problem.  In practice, we have found that the best way to proceed is to express $P_n(r)$, $\rho_n(r)$ in terms of $\Omega_0^2(r)$, $\Omega_2^2(r)$ via equations \eqref{Pn} and \eqref{rhon} and use equation \eqref{Radiative2} for $n=2,4$ to obtain a set of two lengthy, coupled ordinary differential equations for the $\Omega^2_n$. This is a straightforward procedure, but we will not explicitly write out the lengthy equations here, as there is little to be gained beyond the content of equations \eqref{rhobis} and \eqref{Radiative2}.    The $\Omega^2$ equations may be solved, once boundary conditions are specified for $\Omega_0^2$, $\Omega_2^2$ and their first and second derivatives at some radius $r_0$.

\section{Application to the outer radiative zone} \label{sec:application}
In the following analysis, the equilibrium variables $P$, $\rho$ and $g$ have been taken from the solar model of \citet{BahcallSerenelliBasu2005}. The Rosseland opacity $\kappa(\rho, T)$ and its derivatives have been interpolated from the OPAL table for solar composition \citep{IglesiasRogers1996}.

\begin{figure*}
	\centering
	\includegraphics[width=\textwidth, clip=true, trim=0cm 0cm 0cm 0cm]{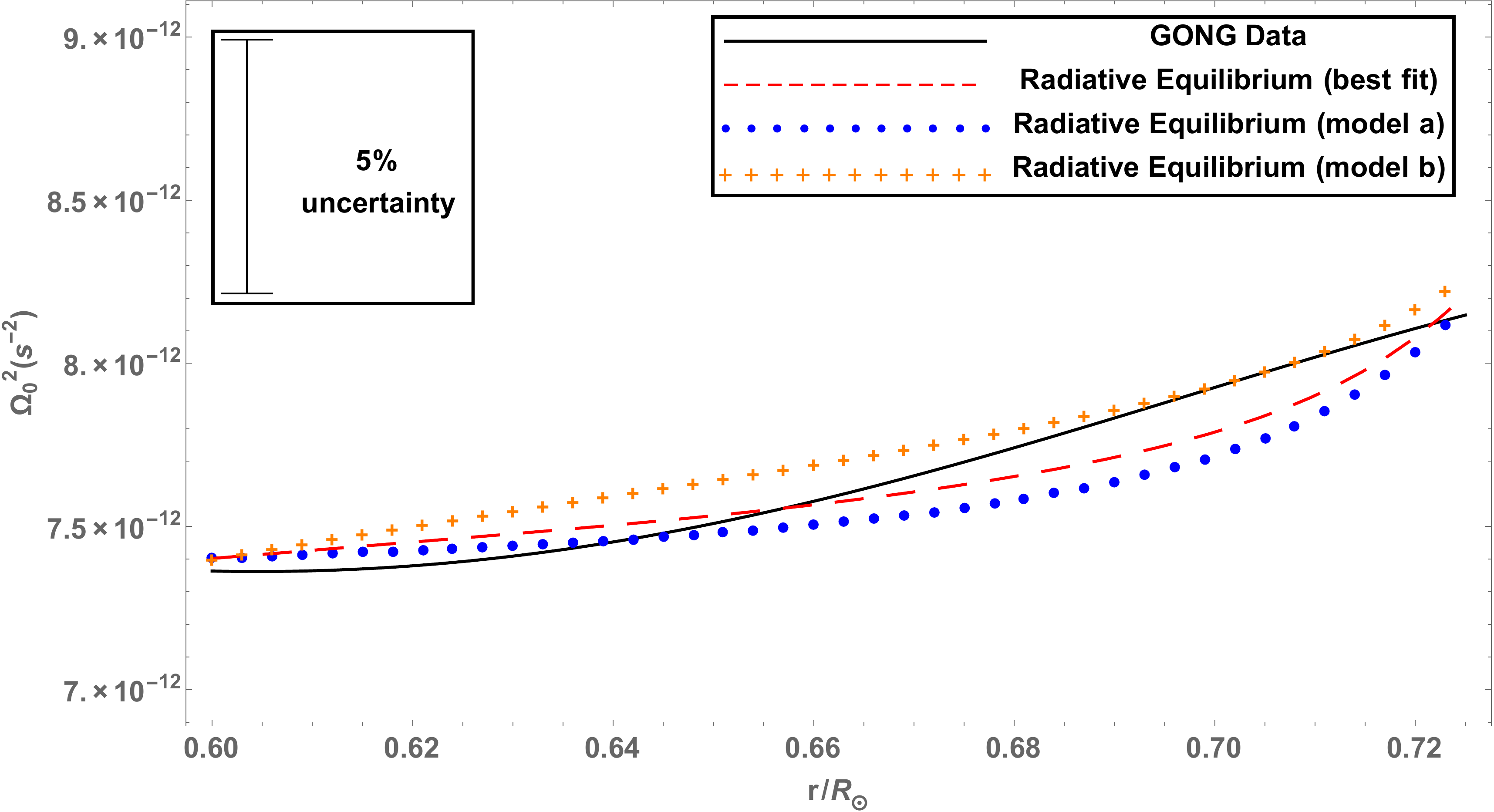}					
	\caption{\label{figOmega0}Angular velocity term $\Omega_0^2(r)$ of equation \eqref{OmegaFit} in the outer part of the radiative zone according to the GONG data (solid line), to the model in radiative equilibrium that best fits the data (dashed line), and to models a (dots) and b (crosses); see description in the text. The boundary conditions at $r_0 = 0.60$ R$_\odot$ are illustrated in table \ref{tab1}. An error bar corresponding to a relative uncertainty of about 5\% at the tachocline is shown, although the actual error is larger.}
\end{figure*}

\begin{figure*}
	\centering
	\includegraphics[width=\textwidth, clip=true, trim=0cm 0cm 0cm 0cm]{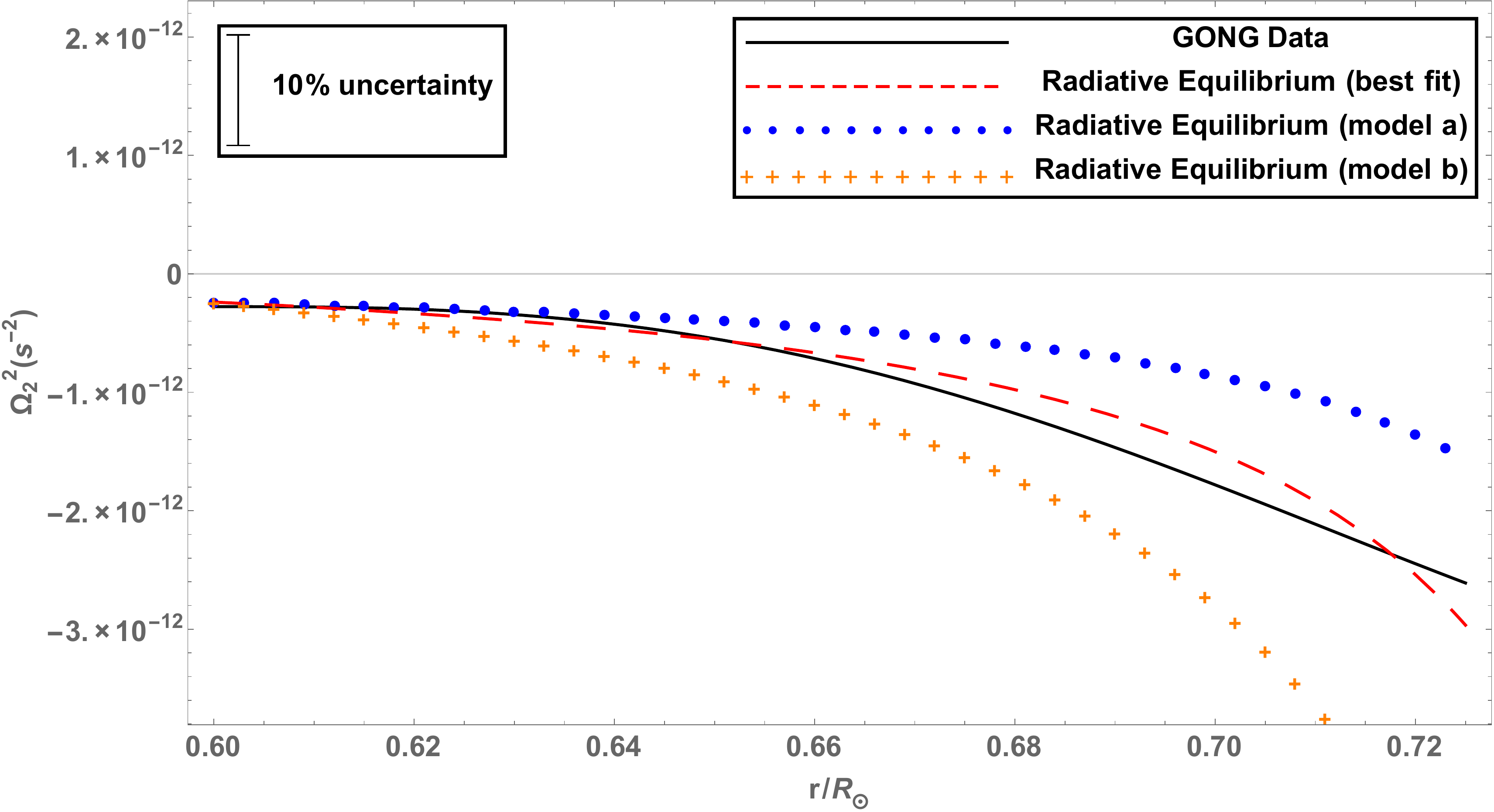}					
	\caption{\label{figOmega2}Angular velocity term $\Omega_2^2(r)$ of equation \eqref{OmegaFit} in the outer part of the radiative zone according to the GONG data (solid line), to the model in radiative equilibrium that best fits the data (dashed line), and to models a (dots) and b (crosses); see description in the text.  The boundary conditions at $r_0 = 0.60$ R$_\odot$ are illustrated in table \ref{tab1}. An error bar corresponding to a relative uncertainty of about 10\% at the tachocline is shown, although the actual error is larger.}
\end{figure*}

Our approach is to solve the coupled $\Omega^2$ equations for comparison with the rotational GONG data.   We set the boundary conditions for $\Omega^2$ and its first and second radial derivatives at $r_0= 0.60 R_\odot$, where the rotation rate is approximately uniform, and integrate outwards.  We adjust the spatial derivatives within the constraint of interior quasi-uniform rotation, and fit the curve to the GONG data. The fit has been executed by performing a $\chi^2$ minimization around an initial, exploratory solution. The resulting boundary conditions imposed at $r_0 = 0.60$ R$_\odot$ are shown in table \ref{tab1}. The angular velocity terms $\Omega_0^2(r)$,  $\Omega_2^2(r)$ of equation \eqref{OmegaFit} for this radiative solution are shown as the dashed lines in figures \ref{figOmega0}, \ref{figOmega2} along with the GONG data.

\begin{table}
\centering
\begin{tabular}{| c | c | c | c |}
\hline
   &  Best fit   &   Model a   & Model b \\ \hline
$\Omega_0^2(r_0)$ & $7.4 \cdot 10^{-12}$ & $7.4 \cdot 10^{-12}$ & $7.4 \cdot 10^{-12}$  \\ \hline
 $d \Omega_0^2(r_0) / dr $ & $3.9 \cdot 10^{-23}$  & $2.0 \cdot 10^{-23}$ & $7.8 \cdot 10^{-23}$ \\ \hline
$d^2 \Omega_0^2(r_0) / dr^2$ & $-4.5 \cdot 10^{-33}$  & $-2.3 \cdot 10^{-33}$ &  $-9.1 \cdot 10^{-33}$  \\ \hline
$\Omega_2^2(r_0)$ & $-2.4 \cdot 10^{-13}$ & $-2.4 \cdot 10^{-13}$ & $-2.4 \cdot 10^{-13}$   \\ \hline
 $d \Omega_2^2(r_0) / dr$ & $-5.9 \cdot 10^{-23}$ & $-3.0 \cdot 10^{-23}$ &  $-1.2 \cdot 10^{-22}$ \\ \hline
$d^2 \Omega_2^2(r_0) / dr^2$ & $-1.2 \cdot 10^{-32}$  & $-0.6 \cdot 10^{-32}$ & $-2.4 \cdot 10^{-32}$ \\ \hline
\end{tabular}
\caption{Boundary conditions imposed at $r_0 = 0.60$ R$_\odot$ for the solutions of figures \ref{figOmega0}, \ref{figOmega2}. The units are in cgs and have been omitted to aid readability.}
\label{tab1}
\end{table}

\begin{figure*}
	\centering
		\subfloat[]{ \includegraphics[width=0.5\textwidth, clip=true, trim=0cm 0cm 0cm 0cm]{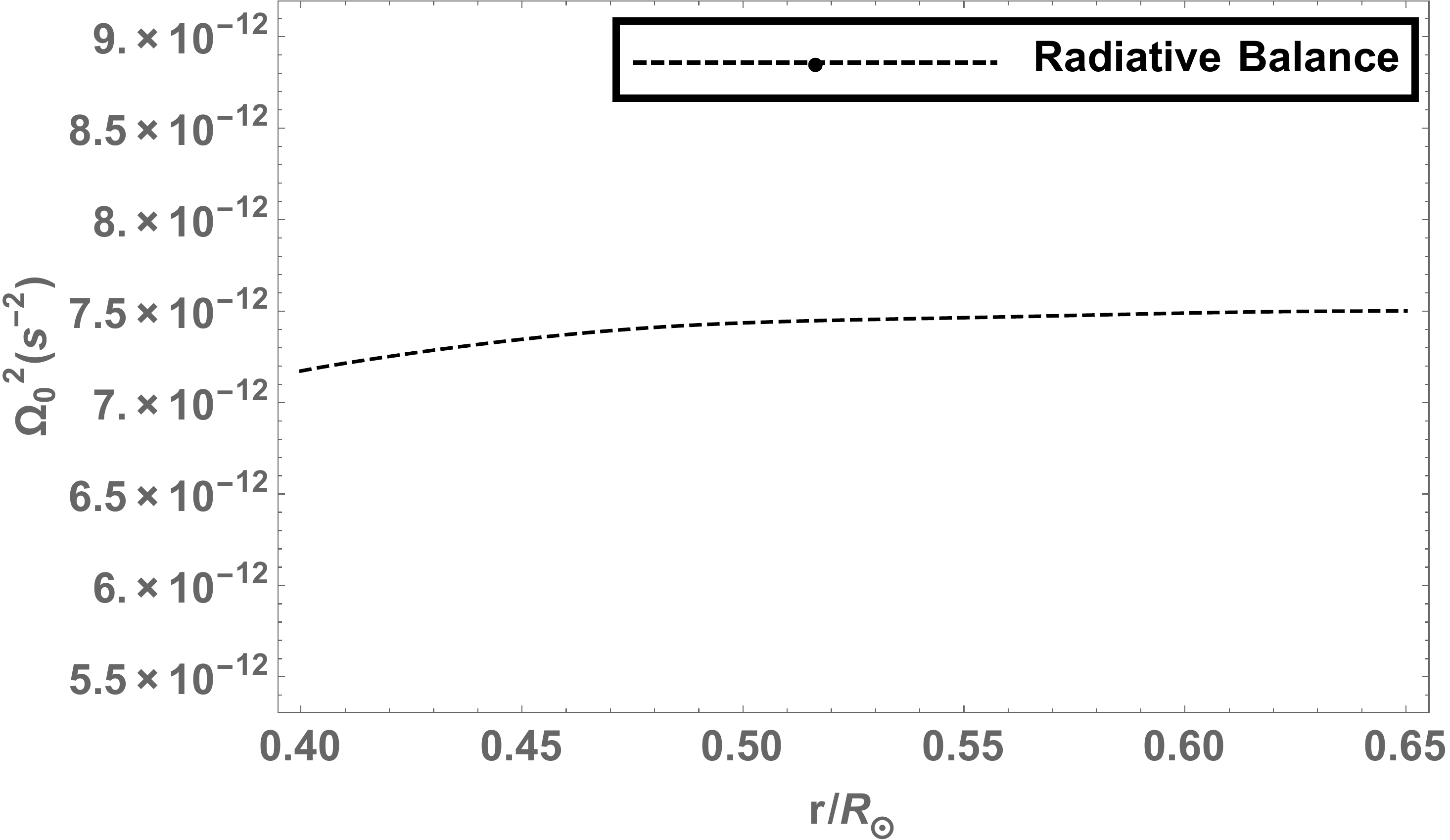} }
		\subfloat[]{ \includegraphics[width=0.5\textwidth, clip=true, trim=0cm 0cm 0cm 0cm]{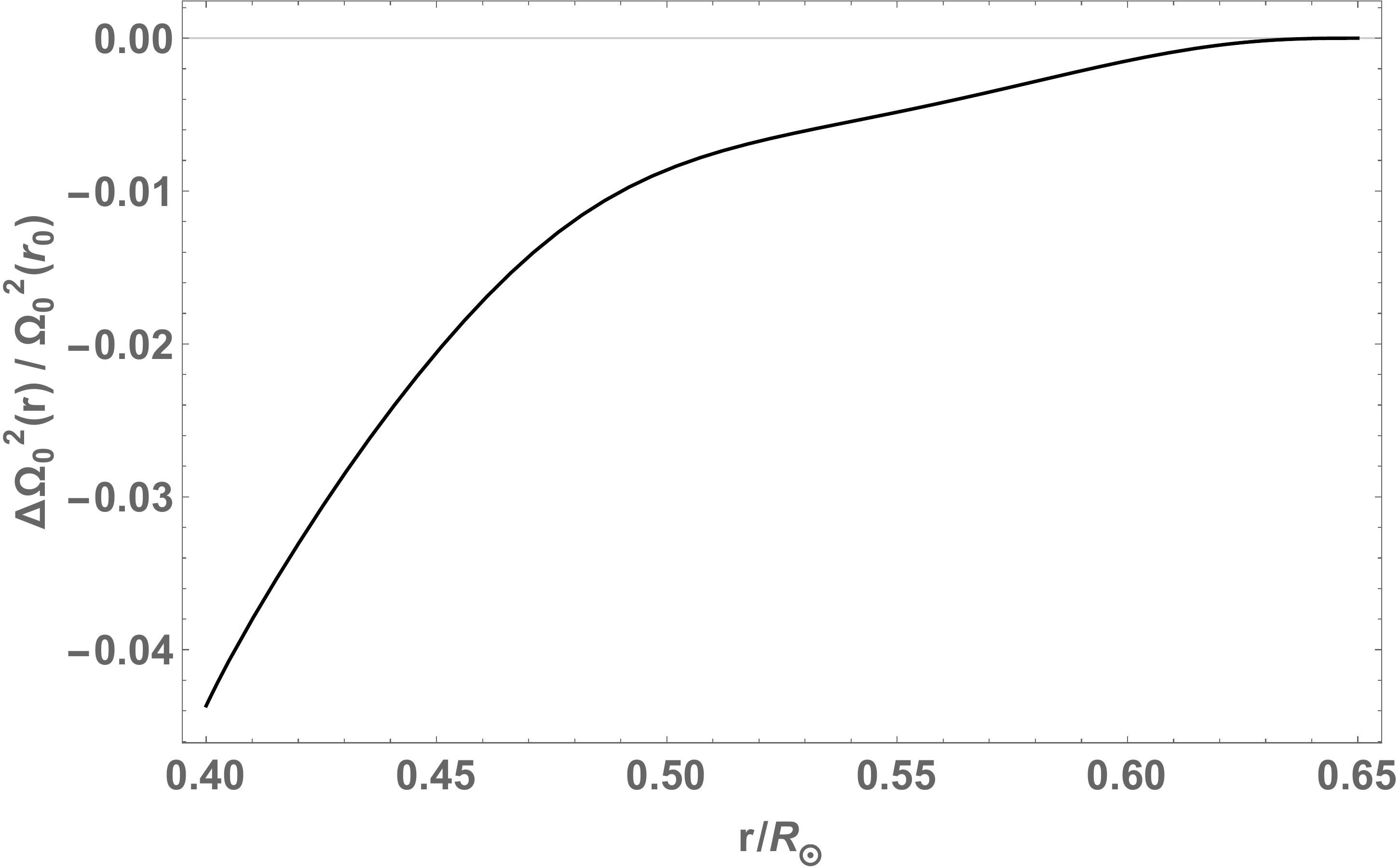} }	
					
	\caption{\label{figOmega3}{\emph a}: Angular velocity $\Omega^2(r)$ in a Sun-like star, obtained imposing boundary conditions of $\Omega_0^2(r_0) = 7.5 \cdot 10^{-12}$ s$^{-2}$ (a typical value of the angular velocity in the Sun), $d \Omega_0^2(r_0) / dr = 0$, $d^2 \Omega_0^2(r_0) / dr^2 = 0$ at $r_0 = 0.65 R_\odot$. \   {\emph b}: Deviation from uniform rotation expressed as $(\Omega_0^2(r) - \Omega_0^2(r_0)) / (\Omega_0^2(r_0))$ with $r_0 = 0.65 R_\odot$.}
\end{figure*}

\emph{Figures \ref{figOmega0}, \ref{figOmega2} indicate that solutions of the model in radiative equilibrium can be found which fit the data well and match closely to a uniformly rotating interior}.  We have shown illustrative 5\% and 10\% error bars for the values of $\Omega_0^2$, $\Omega_2^2$ in figures \ref{figOmega0}, \ref{figOmega2}. It is clear that the model fits well to the data within the expected uncertainties. This result raises the possibility that a very small deviation from uniform rotation might be sufficient to allow the deeper stellar interior  ($R<0.65R_{\odot}$) to be in baroclinic radiative equilibrium.

It should be noted that although the data from helioseismology impose a costraint on the values of $\Omega_0^2(r_0)$, $\Omega_2^2(r_0)$, they are too uncertain to constrain the values of the derivatives of $\Omega_0^2$, $\Omega_2^2$ deep in the radiative zone. It is therefore difficult to compare the derivative boundary conditions we impose with those derived from helioseismology. 

A test of the robustness and sensitivity of the solutions can be performed by noting how they change when the unconstrained boundary conditions $d \Omega_0^2(r_0) / dr$, $d^2 \Omega_0^2(r_0) / dr^2$, $d \Omega_2^2(r_0) / dr$, $d^2 \Omega_2^2(r_0) / dr^2$ are varied. We show in figures \ref{figOmega0}, \ref{figOmega2} the solutions for the case in which the boundary conditions are a factor of 2 smaller (model a) and larger (model b) than those of the best fit.  The respective boundary conditions for each model are shown in table \ref{tab1}. From figures (\ref{figOmega0}) and (\ref{figOmega2}),  it is clear that the validity of the assumption of radiative equilibrium is not acutely sensitive to possible future moderate revisions of the angular velocity profile.

\subsection{Deviation from uniform rotation}

As a preliminary investigation to determine what amount of differential rotation is required for a star to be in radiative equilibrium, the model has been applied to a Solar-like star, i.e. a star with the same pressure, density and temperature profiles as the model by \citet{BahcallSerenelliBasu2005} but a different rotation profile. We integrate towards the deep interior with boundary conditions at $r_0 = 0.65 R_\odot$ locally resembling uniform rotation and using a value of $\Omega(r_0)$ close to the average angular rotation rate of the radiative zone of the Sun. The boundary conditions are shown in table \ref{tab2}.

\begin{table}
\centering
\begin{tabular}{| c | c |}
\hline
$\Omega_0^2(r_0)$ & $7.5 \cdot 10^{-12}$ s$^{-2}$ \\ \hline
 $d \Omega_0^2(r_0) / dr$ & $0$ \\ \hline
$d^2 \Omega_0^2(r_0) / dr^2$ & $0$ \\ \hline
$\Omega_2^2(r_0)$ & $0$ \\ \hline
 $d \Omega_2^2(r_0) / dr$ & $0$ \\ \hline
$d^2 \Omega_2^2(r_0) / dr^2$ & $0$ \\ \hline
\end{tabular}
\caption{Boundary conditions imposed at $r_0 = 0.65$ R$_\odot$ for the solution of figure \ref{figOmega3}.}
\label{tab2}
\end{table}
The model is able to maintain nearly uniform rotation down to a significant depth in the radiative zone by constraining $\Omega_2^2(r)=0$ (see figure \ref{figOmega3}). The solution starts to deviate more strongly (beyond the 1\% level) only at a significant depth (with a deviation of 4\% at $r \sim 0.4$ R$_\odot$).  Were such a profile dynamically stable, the bulk of the radiative zone of a Sun-like star could be in radiative equilibrium and quasi-uniform rotation, without the need for meridional circulation.   The question of stability is further discussed below.

\section{Conclusion}
Meridional circulation is generally seen as a means to maintain thermal energy balance in rotating stars.   However, the purpose of meridional circulation is to maintain uniform or barotropic rotation; if the rotation is baroclinic, it might well be compatible with radiative equilibrium.  Nevertheless, small circulation velocities are usually invoked even in baroclinic flows to maintain thermal balance.  

While solving one problem, meridional circulation seems to create another: starting with the influential study of \citet{SpiegelZahn1992}, circulation-induced diffusive spreading of the tachocline has been viewed as a major problem for understanding solar rotation.   In this paper, we have taken a step back by arguing that since baroclinic solar rotation is a reality, one should investigate what sort of rotation profiles emerge when the constraint of strict radiative equilibrium is applied, without the benefit (or complications) of circulation velocities. We have shown that it is easy to find solutions that fit well to the observed rotation profiles. Indeed, it is not clear that even the deep radiative interior is free of baroclinic rotation at the small level needed to influence radiative balance.   

The most striking feature of the model presented in the current paper is its simplicity.   It is devoid of meridional circulation currents, magnetic fields, viscosity and compositional gradients, effects often invoked to explain the physics of the tachocline \citep{SolarTachocline2007}. It is by far the simplest interpretation that is compatible with the current data from helioseismology.  Interestingly, however, the  problem of time dependence might still implicitly be present in our models, since we have not addressed the all important question of the stability of our baroclinic rotation profiles. The stability of stellar rotation curves is complex because the criteria are significantly affected by such subtlelties as the presence of even weak magnetic fields \citep{Balbus1995}, the thermal diffusion of the displaced fluid elements \citep{GoldreichSchubert}, viscosity \citep{Acheson1978}, and even resistivity \citep{Menou2004}.   One advantage of our formulation is that it is simple enough to lend itself to a rigorous linear stability analysis. If such an analysis shows that the profiles are stable, this would go some way to alleviating the problem of the spreading of the tachocline.  A general three-dimensional, magnetised, nonadiabatic study of the stability of our solutions will be presented in a subsequent paper.

\section*{Acknowledgements}
A. Caleo and W. J. Potter acknowledge support from the University of Oxford. S. A. Balbus acknowledges support from the Royal Society in the form of a Wolfson Research Merit Award. We would like to thank the anonymous referee for an helpful report. A. Caleo would like to thank R. Howe, B. Chaplin, M. Miesch and J. Toomre for useful discussions.

\appendix

\section{Discussion of self-gravity}
In this appendix we estimate the $\delta g_r$, $\delta g_\theta$ terms of equations \eqref{Euler1}, \eqref{Euler2} and discuss the Cowling approximation. For this purpose, it will suffice to consider a star with uniform angular velocity $\Omega$ and to determine the perturbation in the gravitational potential $\delta \Phi$ due to the rotation-induced structural change of the star. 

In this case it is more convenient to expand the pressure and density perturbations in terms of Legendre polynomials $\text{P}_l({\cos \theta})$ rather than in terms of powers of $\cos{\theta}$ as in equations \eqref{PExp}, \eqref{rhoExp}:
\begin{equation} \label{PExp2}
\delta P(r, \theta) =  P_0(r) + P_2(r) \text{P}_2(\mu),
\end{equation} 
\begin{equation} \label{rhoExp2}
\delta \rho(r, \theta) =  \rho_0(r) + \rho_2(r) \text{P}_2(\mu),
\end{equation}
with $\mu = \cos{\theta}$. The $P_0(r)$, $P_2(r)$, $\rho_0(r)$, $\rho_2(r)$ terms of this expansion differ from those of an expansion of the form \eqref{PExp}, \eqref{rhoExp} by a numerical factor.

Plugging expressions \eqref{PExp2}, \eqref{rhoExp2} into equations \eqref{Euler1}, \eqref{Euler2} neglecting the $\delta g_r$, $\delta g_\theta$ terms (hence the perturbation in the gravitational potential will not be self-consistent), and grouping the terms proportional to the same Legendre polynomials in each equation, it is possible to derive $P_2(r)$, $\rho_2(r)$. The procedure is analogous to that leading to equations \eqref{Pn}, \eqref{rhon}. The result for the density perturbation is:
\begin{equation} \label{rho2}
\rho_2 (r) =  \frac{1}{3} \frac{\Omega^2 r^2}{g} \frac{d \rho}{d r}
\end{equation}

The perturbation to the gravitational potential in the star is:
\begin{equation} \label{deltaphi}
\delta \Phi (\bb r) = \int{- \frac{G \delta \rho(\bb r')}{|\bb r - \bb r'|} d^3 \bb r'}.
\end{equation}
Using equation \eqref{rho2} and retaining only the non spherically symmetric part, this is:
\begin{equation} \label{deltaphi2}
\delta \Phi (\bb r) = - \frac{G \Omega^2}{3}  \int{ \frac{r'^2}{g} \frac{d \rho}{d r'} \frac{1}{|\bb r - \bb r'|}  \text{P}_2(\mu') d^3 \bb r'}.
\end{equation}
The integral can be performed by expanding:
\begin{equation} \label{distanceexpansion}
 \frac{1}{|\bb r - \bb r'|} = \sum_{l \geq 0}{\frac{r_<^l}{r_>^{l+1}}\Big(  \text{P}_l(\mu)  \text{P}_l'(\mu') +  \text{Q}_{l}  \Big)},
\end{equation}
where the terms
\begin{equation} \label{distanceexpansion2}
\text{Q}_{l} = \sum_{m \neq 0}{\frac{4 \pi}{2l+1}} \text{Y}^*_{lm}(\theta, \phi)  \text{Y}_{lm}(\theta', \phi'),
\end{equation}
do not contribute to the integral in \eqref{deltaphi2}. Plugging the expansion \eqref{distanceexpansion} in equation \eqref{deltaphi2} the integral is computed by making use of the Legendre polynomials orthogonality properties as:
\begin{equation} \label{deltaphi3}
\delta \Phi = - \frac{4 \pi G \Omega^2}{15} \text{P}_2(\mu) \Big( \frac{1}{r^3} \int_0^r{\frac{r'^6}{g} \frac{d \rho}{dr'} dr'} + r^2 \int_r^{R_\odot}{\frac{r'}{g} \frac{d \rho}{dr'} dr'} \Big).
\end{equation}
By halting the integration at $r' = R_\odot$, we are neglecting the change of the shape of the surface of the star due to the rotation and the effect of the material outside the surface of the non-rotating structure. This is justified as the density in the outer layers, and therefore their contribution to the perturbation in the potential, is low, as verified by our numerical computation of the integrals. 

We determined the potential perturbation $\delta \Phi$ assuming the same Solar background structure as in section \ref{sec:application} and a uniform angular velocity $\Omega^2 = 7.5 \cdot 10^{-12}$ s$^{-2}$. The amplitudes of the resulting gravity perturbations computed from the components of $\del \delta \Phi$ at $r = 0.65$ R$_\odot$ and $\mu = 1$ are:
\begin{equation} \label{deltagr}
|\delta g_r (0.65 R_\odot)| = \Big| \frac{\partial \delta \Phi}{\partial r} \Big| = 1.1 \cdot 10^{-2} \text{cm s}^{-2},
\end{equation}
\begin{equation} \label{deltagtheta}
|\delta g_\theta (0.65 R_\odot)| = \Big| \frac{1}{r} \frac{\partial \delta \Phi}{\partial \theta} \Big| = 1.5 \cdot 10^{-2} \text{cm s}^{-2}.
\end{equation}
The ratios of these terms to the penultimate terms in equations \eqref{Euler1}, \eqref{Euler2} are:
\begin{equation} \label{ratio1}
\Big|\frac{\delta g_r}{g \delta \rho / \rho} \Big| = 1 \%,
\end{equation}
\begin{equation} \label{ratio2}
\Big|\frac{\delta g_\theta}{(1/(r \rho)) (\partial \delta P / \partial \theta)} \Big| = 9 \%.
\end{equation}
We have therefore shown that while the radial self-gravity can safely be neglected, the horizontal one is small but not tiny. The Cowling approximation can be employed in the study of rotating stars, but precision work on their interior structure should be conducted retaining these additional terms. 

\bibliographystyle{mn2e}
\bibliography{References}

\begin{thebibliography}{}

\bibitem[\protect\citeauthoryear{{Acheson}}{{Acheson}}{1978}]{Acheson1978}
{Acheson} D.~J.,  1978, Phil. Trans. R. Soc. Ser. A,, 289, 459

\bibitem[\protect\citeauthoryear{{Antia} \& {Basu}}{{Antia} \&
  {Basu}}{2011}]{Antia2011}
{Antia} H.~M.,  {Basu} S.,  2011, ApJL, 735, L45

\bibitem[\protect\citeauthoryear{{Bahcall}, {Serenelli} \& {Basu}}{{Bahcall}
  et~al.}{2005}]{BahcallSerenelliBasu2005}
{Bahcall} J.~N.,  {Serenelli} A.~M.,    {Basu} S.,  2005, ApJL, 621, L85

\bibitem[\protect\citeauthoryear{{Balbus}}{{Balbus}}{1995}]{Balbus1995}
{Balbus} S.~A.,  1995, ApJ, 453, 380

\bibitem[\protect\citeauthoryear{{Balbus}, {Bonart}, {Latter} \&
  {Weiss}}{{Balbus} et~al.}{2009}]{BalbusBonartLatterWeiss2009}
{Balbus} S.~A.,  {Bonart} J.,  {Latter} H.~N.,    {Weiss} N.~O.,  2009, MNRAS,
  400, 176

\bibitem[\protect\citeauthoryear{{Balbus} \& {Schaan}}{{Balbus} \&
  {Schaan}}{2012}]{BalbusSchaan2012}
{Balbus} S.~A.,  {Schaan} E.,  2012, MNRAS, 426, 1546

\bibitem[\protect\citeauthoryear{{Brun}, {Miesch} \& {Toomre}}{{Brun}
  et~al.}{2011}]{Brun2011}
{Brun} A.~S.,  {Miesch} M.~S.,    {Toomre} J.,  2011, ApJ, 742, 79

\bibitem[\protect\citeauthoryear{{Chaplin}, {Christensen-Dalsgaard},
  {Elsworth}, {Howe}, {Isaak}, {Larsen}, {New}, {Schou}, {Thompson} \&
  {Tomczyk}}{{Chaplin} et~al.}{1999}]{Chaplin1999}
{Chaplin} W.~J.,  {Christensen-Dalsgaard} J.,  {Elsworth} Y.,  {Howe} R.,
  {Isaak} G.~R.,  {Larsen} R.~M.,  {New} R.,  {Schou} J.,  {Thompson} M.~J.,
  {Tomczyk} S.,  1999, MNRAS, 308, 405

\bibitem[\protect\citeauthoryear{{Christensen-Dalsgaard} \&
  {Thompson}}{{Christensen-Dalsgaard} \& {Thompson}}{2007}]{Christensen2007}
{Christensen-Dalsgaard} J.,  {Thompson} M.~J.,  2007, {in Hughes D., Rosner R.,
  Weiss N., eds, The Solar Tachocline. Cambridge Univ. Press, Cambridge, p. 53}

\bibitem[\protect\citeauthoryear{{Garaud}}{{Garaud}}{2007}]{Garaud2007}
{Garaud} P.,  2007, {in Hughes D., Rosner R., Weiss N., eds, The Solar
  Tachocline. Cambridge Univ. Press, Cambridge, p. 147}

\bibitem[\protect\citeauthoryear{{Goldreich} \& {Schubert}}{{Goldreich} \&
  {Schubert}}{1967}]{GoldreichSchubert}
{Goldreich} P.,  {Schubert} G.,  1967, ApJ, 150, 571

\bibitem[\protect\citeauthoryear{{Gough} \& {McIntyre}}{{Gough} \&
  {McIntyre}}{1998}]{GoughMcIntyre1998}
{Gough} D.~O.,  {McIntyre} M.~E.,  1998, Nature, 394, 755

\bibitem[\protect\citeauthoryear{{Hill} \& {et al.}}{{Hill} \& {et
  al.}}{1996}]{GONGData}
{Hill} F.,  {et al.} 1996, Science, 272, 1292

\bibitem[\protect\citeauthoryear{{Howe}}{{Howe}}{2009}]{Howe2009}
{Howe} R.,  2009, Living Rev. Sol. Phys., 6, 1

\bibitem[\protect\citeauthoryear{{Hughes}, {Rosner} \& {Weiss}}{{Hughes}
  et~al.}{2007}]{SolarTachocline2007}
{Hughes} D.~W.,  {Rosner} R.,    {Weiss} N.~O.,  2007, {The Solar Tachocline,
  Cambridge Univ. Press}

\bibitem[\protect\citeauthoryear{{Iglesias} \& {Rogers}}{{Iglesias} \&
  {Rogers}}{1996}]{IglesiasRogers1996}
{Iglesias} C.~A.,  {Rogers} F.~J.,  1996, ApJ, 464, 943

\bibitem[\protect\citeauthoryear{{Menou}, {Balbus} \& {Spruit}}{{Menou}
  et~al.}{2004}]{Menou2004}
{Menou} K.,  {Balbus} S.~A.,    {Spruit} H.~C.,  2004, ApJ, 607, 564

\bibitem[\protect\citeauthoryear{{Miesch}, {Brun} \& {Toomre}}{{Miesch}
  et~al.}{2006}]{Miesch2006}
{Miesch} M.~S.,  {Brun} A.~S.,    {Toomre} J.,  2006, ApJ, 641, 618

\bibitem[\protect\citeauthoryear{{Rogers}}{{Rogers}}{2011}]{Rogers2011}
{Rogers} T.~M.,  2011, ApJ, 733, 12

\bibitem[\protect\citeauthoryear{{Schwarzschild}}{{Schwarzschild}}{1958}]{Schwarzschild1956}
{Schwarzschild} M.,  1958, {Structure and Evolution of the Stars, Dover
  Publications, p. 177}

\bibitem[\protect\citeauthoryear{{Spiegel} \& {Zahn}}{{Spiegel} \&
  {Zahn}}{1992}]{SpiegelZahn1992}
{Spiegel} E.~A.,  {Zahn} J.-P.,  1992, A\&A, 265, 106

\bibitem[\protect\citeauthoryear{{Tassoul}}{{Tassoul}}{2000}]{Tassoul2000}
{Tassoul} J.-L.,  2000, {Stellar Rotation, Cambridge University Press}

\bibitem[\protect\citeauthoryear{{Zahn}}{{Zahn}}{2007}]{Zahn2007}
{Zahn} J.-P.,  2007, {in Hughes D., Rosner R., Weiss N., eds, The Solar
  Tachocline. Cambridge Univ. Press, Cambridge, p. 89}

\end{thebibliography}
\bibdata{References}

\label{lastpage}

\end{document}